\documentclass{article}

\usepackage{cite}
\usepackage{graphicx}
\usepackage{nicefrac}
\usepackage{amsmath,amsfonts,amssymb}

\begin{document}

\title{Angle-dependent reflectivity of twill-weave carbon fibre reinforced polymer for millimetre waves}

\author{G. Artner, E. Z\"ochmann, S. Pratschner, M. Lerch, \\ M. Rupp and C. F. Mecklenbr\"auker}

\maketitle

\section*{Abstract}

Twill-weave carbon fibre reinforced polymer is measured as reflector material for electromagnetic waves at $60$\,GHz.
The reflectivity at millimetre wavelength shows a predominant direction, which coincides with the diagonal pattern of the twill's top layer.
In the remaining directions the investigated 2/2 twill-weave composite's reflectivity is almost isotropic.

\section{Introduction}

Carbon fibre reinforced polymer (CFRP) are composites, with carbon fibres embedded in a polymer matrix.
Nowadays, they are widely used in the construction of lightweight vehicles and antennas.
Typical applications of CFRP are reflectors for spacecraft antennas~\cite{Keen1975}.
A millimetre wave (mmW) antenna with a CFRP reflector for radar applications is presented in \cite{Futatsumori2013,Futatsumori2014}.

Measurements of CFRP reflectivity and conductivity consider mostly unidirectional fibres and  few material orientations~\cite{Klooster2003,Riley2015,Artner2017}.
High precision reflectivity measurements in the range of $110 \textrm{-} 200$\,GHz of unidirectional CFRP measured in the fibre direction and perpendicular to the fibre direction are conducted with a Fabry-Perot resonator in \cite{Klooster2003}. There, the reflector is mounted in fixed orientations and material orientation is not considered.
Riley et al.  present measurements of the radar cross section of twill-weave CFRP in \cite{Riley2015}, but only in a mono-static set-up and sample orientation was not considered.
Artner et al. estimate the conductivity of twill-CFRP from wave-guide measurements at $4 \textrm{-} 6$\,GHz with the Nicolson-Ross-Weir method in steps of $10^\circ$~\cite{Artner2017}.

Contribution: In this letter, the anisotropy of twill-weave CFRP reflectivity at millimeter wavelength is observed in a bi-static set-up.
Measurements are performed with millimetre waves at $60$\,GHz.
Several widely used CFRP fabrics were investigated: unidirectional fibres, shredded fibres, plain weave and twill weave.
A sharp angular dependency in reflectivity of twill-weave CFRP is found around angles, where the direction of the twill pattern aligns with the incident wave's electric field vector.
The sharp angular dependency is only found in bi-static set-ups and only for the twill-weave's reflectivity.
Other investigated CFRPs (plain-weave, unidirectional, shredded-fibres) are omitted, as measurements do not show the  observed effect.
To the best of the authors' knowledge this effect has not been measured or described before.

\section{Twill-weave CFRP}

\begin{figure}

\centering
\includegraphics[width=80mm]{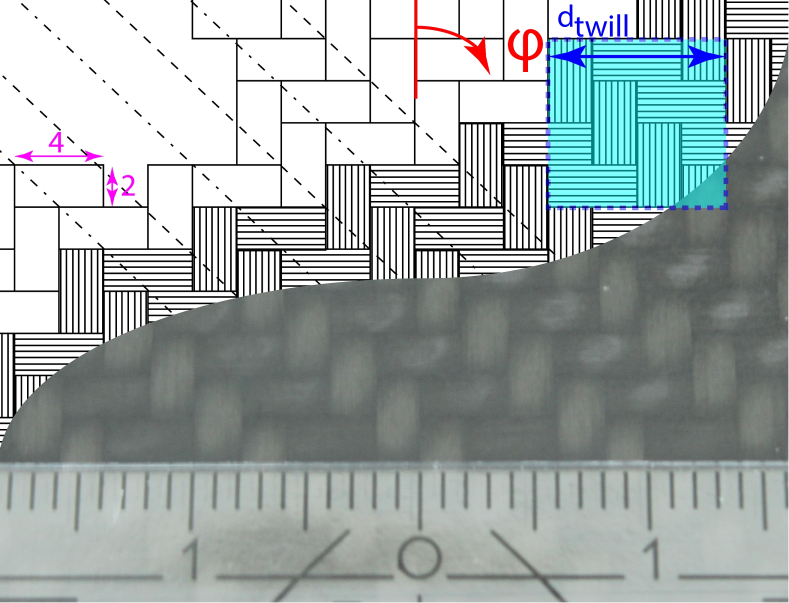}
\caption{Photograph of the measured $\nicefrac{2}{2}$ twill-weave CFRP. Ruler dimensions are in centimetre. The photograph is sketched with various degrees of abstractions: It reveals fibre strand orientation, twill-weave pattern and the distinctive diagonal twill-direction (dashed lines). The blue dotted square marks the twill weave pattern.}
\label{fig:twill}

\end{figure}

The material under test (MUT) is a CFRP laminate of $1.6$\,mm total thickness, produced from $\nicefrac{2}{2}$ twill-weave plies with the press method.
A circular disk with a diameter of $d_{\textrm{MUT}} = 280$\,mm was cut by a water-jet.
Sheets are commercially available. 
A photograph of the MUT is displayed in Fig. \ref{fig:twill}.
Sketches are superimposed, which reveal fibre-, strand- and twill-pattern-directions.
On the surface, the fibre strands  have a visible size of about $2 \times 4$\,mm$^2$, which is within the same order of magnitude as the free space wavelength at $60$\,GHz of $\lambda \approx 5$\,mm.
The blue dotted square in Fig.~\ref{fig:twill} shows the twill-weave pattern (\emph{Bindungspatrone}). 
This quadratic pattern of size $d_{\rm twill} \times d_{\rm twill}$ has a side length of $8 $\,mm.  
Theoretically the twill-weave CFRP is an example of a finitely conducting bigrating with $\nicefrac{\lambda}{d_{\rm twill}} \approx \nicefrac{5}{8}=0.625$, see~\cite{Petit1980} Section 7.6, pp. 258ff.
The lower layers' geometry is unimportant due to the electromagnetic waves' low penetration depth resulting from the skin effect.

\section{Measurement set-up}
\begin{figure}[h]
\centering{\includegraphics[height=35mm]{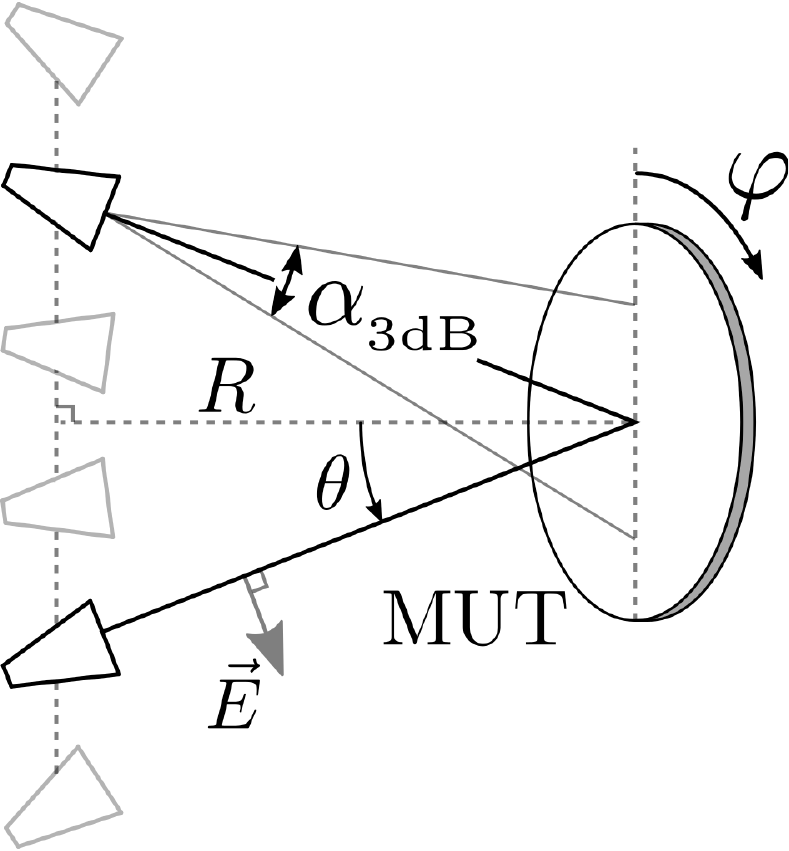}\quad~\includegraphics[height=35mm]{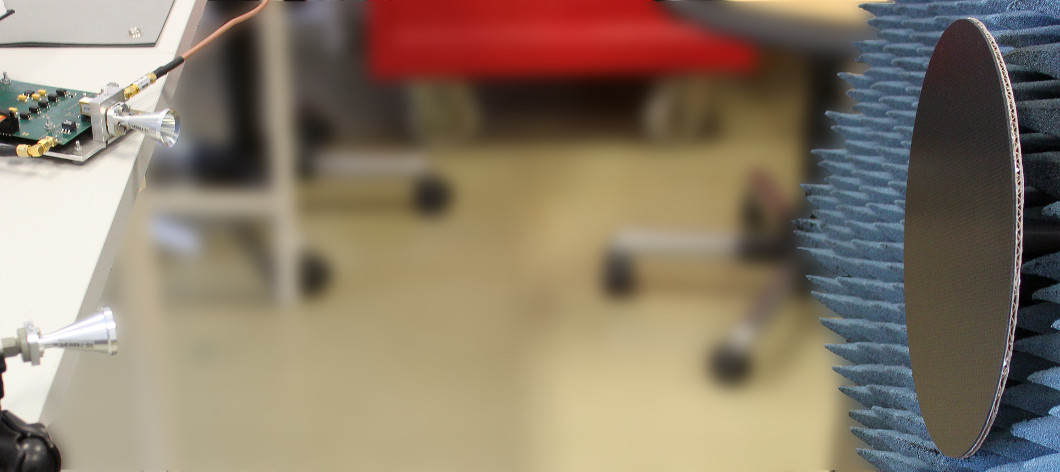}}
\caption{Geometry of the measurement set-up, top view (left). Photograph of the two horn antennas and the MUT in front of absorbers (right).}
\label{fig:setup}
\end{figure}

Two identical, horizontally polarised conical horn antennas are used in a bi-static set-up as shown in Fig.~\ref{fig:setup}. 
This polarisation corresponds to the transversal magnetic (TM) reflection case. 
Electromagnetic waves with a center frequency of $60\,$GHz are transmitted by one antenna, reflected by the MUT and received by a second antenna.
The baseband transmit signal is a  $2\,$GHz wide multi-tone waveform with Newman phases to ensure a low crest factor~\cite{Newman1965}.
It is generated with an arbitrary waveform generator and up-converted with an external mmW-module.
At the receiver side, the $60$\,GHz signal is directly measured with a spectrum analyser.
The received power over the whole bandwidth is recorded and subsequently averaged.

The MUT disk is placed on a rotary plate to adjust the turn angle $\varphi$ automatically in steps of $1^\circ$.
This allows automated measurement of material anisotropy, by changing the MUT's alignment towards the incident polarization.
The reflection angle $\theta$ is adjusted by changing the distance between the antennas.
The normal distance $R$ is kept constant, due to the mounting of the antennas as shown in Fig.~\ref{fig:setup}.

For each reflection angle $\theta$, the antennas are aligned to point towards the MUT's center.
The horn antennas have an aperture of $27.41\,$mm and a 3\,dB opening angle of $\alpha_{3{\rm dB}}=18^\circ$.
The distance $R$ was chosen to be $55$\,cm, such that the MUT is in the far-field and most of the energy is focused on the MUT.
Absorbers are placed around the MUT to eliminate undesired reflections.

\section{Measurement results}

\begin{figure}
\centering{\includegraphics[width=1\textwidth]{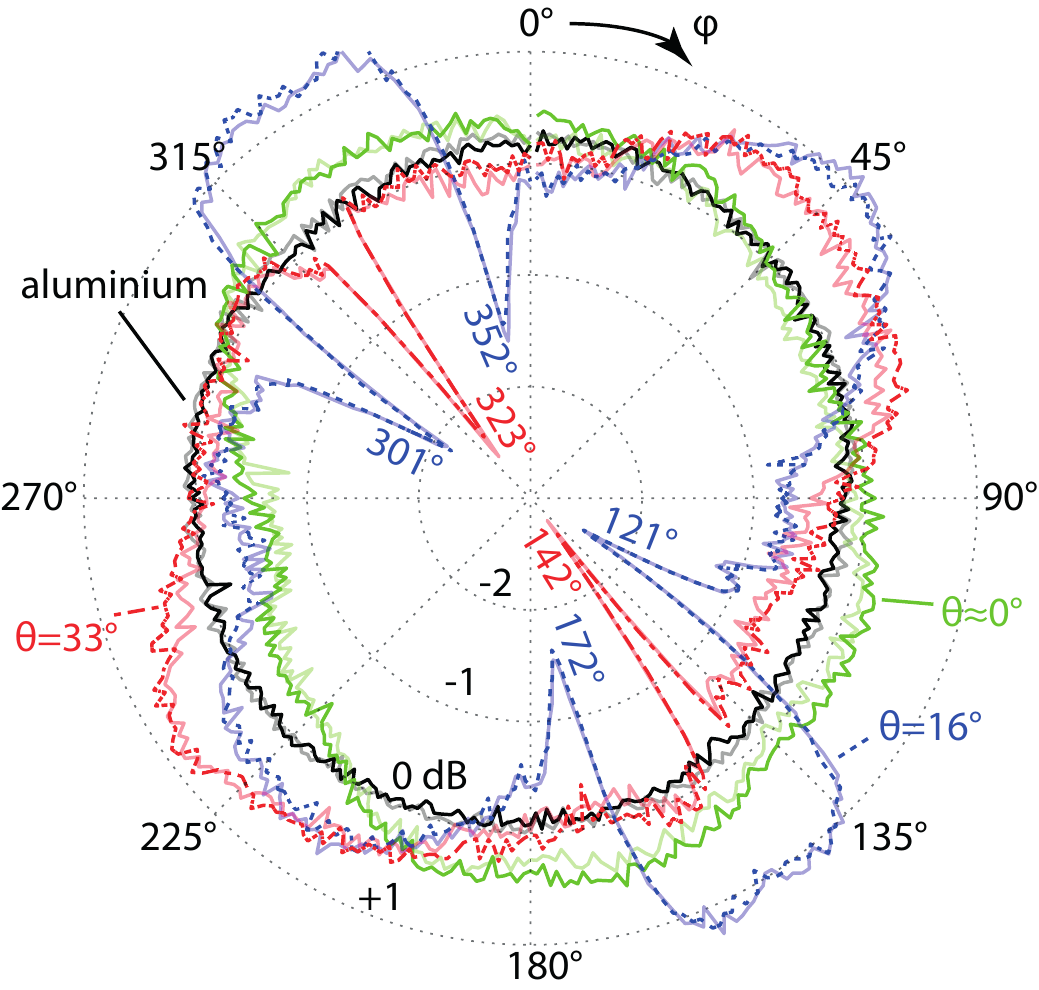}}
\caption{Normalised measured reflected power of twill-CFRP for two full turns of the MUT. The second turn is displayed opaque.}
 \label{fig:results}
\end{figure}

We observe a sharp change in reflectivity at angles corresponding to the diagonal twill pattern on the top layer.
Measured results of the twill-CFRP's reflectivity are presented in Fig.~\ref{fig:results}.
Due to the constant distance $R$, the average line-of-sight distance to the MUT depends on the reflection angle $\theta$.
To allow comparison of the angular dependency, receive power is normalised, i.e., $P(\varphi,\theta) / \mathbb{E}_\varphi \left\{ P(\varphi,\theta) \right\}$, where $\mathbb{E}_\varphi$ denotes the expectation operator with respect to $\varphi$. 

The estimated temperature coefficient of our measurement set-up is $-0.14$\,dB/K.
During our measurements, the temperature varied by $\pm 0.7$\,K.
The receive power is corrected by the estimated temperature coefficient.

Measurements are performed for $\theta \in \lbrace 0^\circ,16^\circ,33^\circ \rbrace$. 
At $\theta\approx 0^\circ$ the antennas are placed above each other. 
For each elevation angle $\theta$ the MUT is rotated two full turns in azimuth $\varphi$.
Both curves are shown in Fig.~\ref{fig:results} in order to visualize the influence of MUT rotation precision.

Aluminium's measured isotropic reflectivity is shown as comparison to illustrate estimation inaccuracies. 

The curves show reflectivity dependency on the MUT orientation $\varphi$.
A $180^\circ$ periodicity is evident due to the MUT twill-weave.

Sharp changes in reflectivity are observed at certain angles of MUT orientation $\varphi$, close to the diagonal twill-direction.
Depending on reflection angle $\theta$, one or two notches appear.
The notches are not observed for the mono-static case, i.e., $\theta \approx 0^\circ$.
The notches appear around the diagonal twill-direction and also have a periodicity of $180^\circ$.
We therefore conclude, that these sharp angular changes in reflectivity are caused by the CFRP twill pattern.
Surprisingly, no significant influence is found in the directions of warp and weft $\varphi \in \lbrace 0^\circ,90^\circ,180^\circ,270^\circ \rbrace$, where fibre directions coincide with field vectors.

\section{Conclusion}
Twill-CFRP's electromagnetic reflectivity is anisotropic at millimetre wavelengths.
Sharp notches in reflectivity are observed, that appear around the predominant direction of the diagonal twill pattern.
These notches do not appear for mono-static set-ups.

In the directions of warp and weft $\varphi \in \lbrace 0^\circ,90^\circ,180^\circ,270^\circ \rbrace$ no significant influence on reflectivity is observed.

A quantitative numerical electromagnetic simulation for the twill-weave CFRP is still open as well as full elevation scans of reflected power.

\vspace{1cm}

\noindent G. Artner, E. Z\"ochmann, S. Pratschner, M. Lerch, M. Rupp and C.~F. Mecklenbr\"auker (\textit{Institute of Telecommunications, Technische Universit\"at Wien, 1040 Vienna, Austria})

\noindent E-mail: gerald.artner@nt.tuwien.ac.at

\end{document}